\documentclass[reprint,twocolumn, aps, prb]{revtex4-1}

\usepackage{graphicx,psfrag}
\usepackage{epstopdf}
\usepackage{amsmath}
\usepackage{bm}
\usepackage{amssymb}
\usepackage{subfigure}

\begin{document}


\title{Tomographic imaging of inhomogeneous non-local media using fractional order models}

\author{Salvatore Buonocore}
 \affiliation{University of Notre Dame, Notre Dame, Indiana, 46556.}
\author{Fabio Semperlotti}%
 \email{fsemperl@purdue.edu}
\affiliation{ 
Purdue University, Ray W. Herrick Laboratories, West Lafayette, Indiana, 47907
}%

\date{\today}

\begin{abstract}
We investigate a generalized tomographic imaging framework applicable to a class of inhomogeneous media characterized by non-local diffusive energy transport. Under these conditions, the transport mechanism is well described by fractional-order continuum models capable of capturing anomalous diffusion that would otherwise remain undetected when using traditional integer-order models. Although the underlying idea of the proposed framework is applicable to any transport mechanism, the case of fractional heat conduction is presented as a specific example to illustrate the methodology. By using numerical simulations, we show how complex inhomogeneous media involving non-local transport, can be successfully imaged if fractional order models are used. In particular, results will show that by properly recognizing and accounting for the fractional character of the host medium not only allows achieving increased resolution but, in case of strong and spatially distributed non-locality, it represents the only viable approach to achieve a successful reconstruction. 
\end{abstract}

\maketitle

\section{INTRODUCTION}
Tomography, or tomographic imaging, is a particular class of inverse problems that enables reconstructing an image of the internal structure of a body (or of specific material properties) without requiring intrusive access to it. Depending on the specific method, different types of penetrating waves or field transport phenomena can be selected to probe the body. Typical examples include electrostatic and electromagnetic waves (employed in Electrical Impedance Tomography-EIT and Magnetic Resonance Imaging-MRI, respectively) \cite{holder2004electrical}, acoustic waves (acoustic tomography) \cite{munk2009ocean}, and thermal fields (thermography) \cite{vavilov2010thermal}. In the past few decades, tomographic techniques have found many engineering applications including biomedics, atmospheric science, geology, oceanography, plasma physics, materials science, astrophysics, and acoustics 
\cite{ammari2008introduction,norton1989tomographic,bayford2006bioimpedance}. More recently, tomography was also successfully applied to structural health monitoring and non-destructive testing for the inspection and characterization of structures and materials \cite{vavilov2010thermal,verstrynge2014novel,swiderski2009characterization, SEMP1,SEMP2,SEMP3,SEMP4}.

As many inverse problems, tomographic methods are heavily dependent on the accuracy of the underlying mathematical model used to predict the response of the medium. Typically, inverse methodologies fit, in a least-squares sense, a mathematical model to experimentally measured data by iteratively tuning the model parameters
\cite{holder2004electrical,ozisik2000inverse,bendsoe2004topology}. It is evident that the ability to reconstruct accurate images and to improve the resolution is strongly dependent on the ability of the model to capture and properly simulate the underlying physical response of the system.
Typical applications of tomographic methods deal with domains of analysis that are highly inhomogeneous (e.g. including geometric or material inhomogeneities), have complex material properties (e.g. viscoelastic materials), and are eventually characterized by hybrid and non-local transport behavior. If not properly accounted for, these properties limit both the sensitivity and resolution of the methodology and can even prevent entirely the possibility to reconstruct a meaningful image.

To-date, several numerical and experimental studies have demonstrated that field transport processes in complex inhomogeneous media (e.g. random and/or porous materials) can occur according to hybrid modalities \cite{mainardi1996fractional,mainardi1996fundamental,mainardi1994special,chen2003modified,chen2004fractional,mainardi2010seismic} and anomalous behavior\cite{benson2000application,benson2001fractional,cushman2000fractional,fomin2005effect}. The hybrid transport typically involves a combination of propagating and diffusive mechanisms\cite{asatryan2003diffusion,sheng2006introduction,sheng1990scattering,barthelemy2008levy,bayer1993weak}.

The diffusion processes can be either standard, that is characterized by a typical Gaussian distribution of the field quantities, or anomalous where instead an $\alpha$-stable distribution emerges. In general, diffusive processes are characterized based on the functional relationship that describes the evolution of the variance of the particle displacement with time. This relationship is expressed as $\left \langle x \right \rangle^2\sim t^\gamma$ where the normal diffusion is characterized by a linear scaling $\gamma=1$, while the anomalous diffusion occurs for $\gamma \neq 1$. In particular, processes described by $\gamma>1$ correspond to diffusion phenomena that evolve in time faster than the standard (Gaussian) process and are therefore denoted as \textit{superdiffusive}. On the contrary, processes that are characterized by $0<\gamma<1$, are denoted as \textit{subdiffusive}, because the variance increases more slowly than what predicted by the classical (Gaussian) diffusion model. Subdiffusive processes can be described at a microscopic level by continuous time random walk (CTRW) models employing a probability density function for the waiting time of a particle jump characterized by heavy tailed distribution \cite{metzler2000random,bouchaud1990anomalous}. On the other hand, superdiffusive phenomena can be described by L{\'e}vy flights or L{\'e}vy walks where the probability density function of the step length is a heavy-tailed distribution (that emerges as a consequence of the long-range interactions among particles). 

These unconventional transport mechanisms are typically driven by non-local effects, such as long-range interactions \cite{tarasov2006fractional,tarasov2014lattice}, memory and hereditary mechanisms\cite{tarasov2008conservation,tarasov2015non}, and have been shown to be well described by fractional order models. At macroscopic scales, the CTRW subdiffusive processes can be mapped to a time-fractional diffusion equation while the superdiffusive processes map into a space-fractional diffusion equation. Examples of these unusual transport behavior can be observed in different media and fields of application including, but not limited to, porous soils \cite{
caputo1971new,caputo1971linear,bagley1986fractional,meral2010fractional,
mainardi2010fractional}, heterogeneous aquifers \cite{benson2000application,benson2013fractional}, and underground fluid flow \cite{benson2002fractional,fomin2005effect}. Meerschaert et al. \cite{benson2002fractional,schumer2001eulerian} performed an extensive research campaign focused on the transport of contaminants in heterogeneous soils and porous materials and showed that the underlying advective-diffusive behavior exhibits indeed a fractional nature. Other studies \cite{barthelemy2008levy,cobus2016anderson,
burresi2012weak,stutzer2013observation,
bouchaud1990anomalous,fellah2000transient} have highlighted the occurrence of regular and anomalous diffusion mechanisms in the propagation of electromagnetic and acoustic waves in random and periodic materials. Propagation of light through fog \cite{belin2008display} or murky media \cite{zevallos2005time}, of sound through a forest or a urban environment
\cite{chobeau2014modeling,hornikx2016ten,kang2000sound}, of seismic waves through the ground \cite{fellah2003measuring,fellah2003solution} are all examples of practical applications where hybrid and non-local transport mechanisms can take place. In all these situations, fractional calculus was shown to be a powerful tool to properly capture and simulate complex transport mechanisms in continuum media.

From the discussion above, it appears that the integration of fractional order models within tomographic methods could improve the imaging performance and even allow capturing details and inhomogeneities that would otherwise go undetected when using conventional imaging approaches based on integer order models.

In this paper, we present a numerical investigation into the application of fractional order continuum models to tomographic imaging techniques.
A few applications of fractional calculus to inverse problems have already been explored in the literature. Cheng et al. \cite{cheng2009uniqueness} successfully solved a one-dimensional fractional diffusion inverse problem to determine the order of the temporal fractional derivative and the spatially varying diffusion coefficients. 
Miller et al. \cite{miller2013coefficient} formulated a parameter identification problem based on the fractional diffusion equation. Battaglia et al. \cite{battaglia2001solving} proposed the solution of an inverse heat conduction problem for parameter estimation based on non-integer forward models. The precision of the inverse procedure proposed in their method heavily depended on the accuracy of the identified model that had to be determined by experiments. Murio \cite{murio2008time,murio2007stable} presented the numerical solution of an ill-posed problem consisting in determining analytical functions of the boundary temperatures and the heat fluxes based on transient temperature measurements at some internal point of a 1D conductor. Jin and Rundell \cite{jin2015tutorial} investigated the degree of ill-posedness of a series of theoretical and numerical inverse problems based on fractional differential equations involving the Caputo definition in both time and space. They found out that the fractional character of the operator can either improve or worsen the conditioning of the inverse problem based on the type of input data and quantities to be reconstructed. Kirane et al. \cite{kirane2013inverse} solved a 2D source reconstruction problem for a time fractional diffusion equation using biorthogonal sets of functions.

In this study, we explore the capabilities and performance of fractional tomography applied to two-dimensional domains. More specifically, we refer to fractional tomography as a tomographic approach relying on fractional-order continuum models for the solution of the forward problem. The use of these models within a tomographic imaging framework is expected to impact the reconstruction performance at multiple levels:

\begin{enumerate}
	
\item It was previously discussed that anomalous diffusion processes, such as those arising in imaging of highly aberrating and scattering media, are associated with non-Gaussian distributions of the field quantities having noticeable heavy-tails. Note that a distribution of a random variable is said to be heavy-tailed if the moment generating function of the distribution function always diverges \cite{nolan2003stable}. Typically, the content of the tails is discarded because not considered as a primary source of information for the imaging process. However, while this could be a reasonable assumption when the domain is dominated by classical diffusion (in this case the tails of the Gaussian distribution have a low content of information), in presence of anomalous diffusion the tails contain a non-negligible amount of information about the interior structure of the domain to be imaged. Hence, neglecting the heavy tails results in considerable and irremediable loss of information.
	
\item Recognizing and exploiting the fractional nature of the host medium and the information contained in the heavy tails can represent a turning point in the development of accurate imaging technologies capable of sensing through highly inhomogeneous and scattering media.

\item In order to extract meaningful information from the non-Gaussian distribution of the field quantities, the underlying mathematical model should be able to simulate these hybrid and anomalous physical mechanisms and to relate them to the physical parameters of the medium.
 
\end{enumerate}
In order to clearly illustrate the methodology and without limiting the generality of our approach, we focus the following study on diffusive transport processes. An example of a classical tomographic method that falls under this category is Thermal Tomography (TT) \cite{swiderski2009characterization,
vavilov2010thermal,vavilov1992some,kline2003new}, where diffusive heat transfer is used as a probing mechanism to sense the medium. Note that the results presented below could be easily generalized to other approaches whose transport mechanism can be effectively described by fractional models.

\section{Fractional Tomography: imaging based on fractional order models}

Tomographic imaging techniques are usually formulated as iterative inverse problems in which measured data are fit, in a least-squares sense, to a mathematical model that simulates the response of the host medium \cite{holder2004electrical}. Although many inversion methodologies have been presented over the years, the most common approach to solve complex geometries is based on iterative numerical techniques. These methods typically consist of two main steps: the \textit{forward} and the \textit{inverse} problem. The forward problem simulates the response of the system assuming that the excitation conditions and the system parameters (e.g. material properties, etc.) are known. The inverse problem relies on an optimization approach designed to minimize the least-square error between the measured and the predicted response at selected measurement points. The system parameters, or a subset of them, are typically selected as design variables. 

In this work, we use fractional order models in the forward problem in order to describe the heat transport process at the basis of the thermographic technique. We will show that the introduction of fractional models has two main effects on the general formulation of the tomographic problem: 1) if the behavior of the real physical system is fractional, the model will capture information that would otherwise go undetected when using an integer order model, however 2) if the behavior is not fractional but the system is simply inhomogeneous, the fractional formulation will provide an enlarged parameter space (which now includes the order of the differential operator) in which the reconstruction can be performed.

From a general perspective, thermal tomography reconstructs the internal (thermal) properties of a solid medium based on the measurements of boundary (or surface) temperature fields. The data acquisition procedure is performed by heating
the sample at a specific location and measuring the resulting surface temperature at multiple locations. The same procedure is repeated for several source locations in order to improve the conditioning of the inverse problem and increase the accuracy of the reconstruction.
In the numerical analyses performed in this study, the stationary heat sources were placed near the boundaries as shown in Fig.~\ref{Fig0}. The response of the system was estimated in terms of absolute temperature measurements performed at the sensor locations (Fig.~\ref{Fig0}). The thermographic technique is typically formulated as an optimization process based on iterative solution of the forward and inverse problem. The thermal properties, such as the conductivity or the heat capacity, are typically used as design variables.

In the forward problem, the thermal properties are assumed known (either due to an initial guess or to values available from the previous iteration) so that the mathematical model allows estimating the thermal field in the entire domain. In the inverse problem, the measured temperature values at selected locations are compared with numerical predictions in order to find an updated distribution of thermal properties that minimizes the least-square error. The reconstruction of different thermal properties is an indicator either of material inhomogeneities or of possible defects, depending on the specific application of the tomographic technique. 


\subsubsection{Forward problem: fractional heat transport}

The fractional heat diffusion process in a 2D domain is described by the following fractional partial differential equation \cite{meerschaert2006finite}:

\begin{eqnarray} \label{FPDE}
\begin{split}
\frac{\partial^{\gamma}T(x,y,t)}{\partial t^{\gamma}}= c_{1_+}(x,y)\cdot \frac{\partial^{\alpha}T(x,y,t)}{\partial_+ x^{\alpha}}+ \\
c_{1_-}(x,y)\cdot \frac{\partial^{\alpha}T(x,y,t)}{\partial_- x^{\alpha}}+ c_{2_+}(x,y)\cdot \frac{\partial^{\beta}T(x,y,t)}{\partial_+ y^{\beta}}+ \\
c_{2_-}(x,y)\cdot \frac{\partial^{\beta}T(x,y,t)}{\partial_- y^{\beta}}+s(x,y,t)
\end{split}
\end{eqnarray}

where $T(x,y,t)$ is the temperature field at time $t$ in the finite domain $ a<x<b$, $c<y<d$, the terms $c_{1_\pm}(x,y)$, $c_{2_\pm}(x,y)$ are spatially dependent functions representing the thermal diffusivity and are associated with the left- and right-handed fractional operators as indicated by the directional sign notation. $s(x,y,t)$ represents the heat source.

In the following simulations, we will assume that $c_{1_+}(x,y) = c_{1_-}(x,y)= c_{2_+}(x,y)= c_{2_-}(x,y)=c(x,y)$ which means that we consider a symmetric non-local diffusion process. The thermal diffusivity in Eq.(\ref{FPDE}) has dimension $L^{\alpha}t^{-\gamma}$\cite{mainardi2010fractional}. The notation $\partial_\pm$ in Eq. (\ref{FPDE}) refers to the left- and right-handed Riemann-Liouville fractional derivative defined as:

\begin{eqnarray} \label{RLL}
\begin{split}
(D^{\alpha}_{a+} f)(x)=\frac{\partial^{\alpha}f(x)}{\partial_+x^\alpha}=\frac{1}{\Gamma(n-\alpha)}\frac{d^n}{dx^n} \int_{a}^{x}\frac{f(\xi) d\xi}{(x-\xi)^ {\alpha+1-n}} \\(D^{\alpha}_{b-} f)(x)=\frac{\partial^{\alpha}f(x)}{\partial_-x^\alpha}=\frac{1}{\Gamma(n-\alpha)}\frac{d^n}{dx^n} \int_{x}^{b}\frac{f(\xi) d\xi}{(\xi-x)^ {\alpha+1-n}}
\end{split}
\end{eqnarray}

\noindent where $\Gamma$ is the Euler gamma function and $n$ is an integer such that $n-1 \leq \alpha \leq n $.

\noindent Equation (\ref{FPDE}) represents a general form of the fractional wave-diffusion equation capable of representing several forms of transport mechanisms \cite{mainardi2007fundamental}. When $0 <\gamma \leq 2$ and $\alpha=\beta=2$, the equation describes a time-fractional process having diffusion-like characteristics when $0 < \gamma \leq 1$ and wave-like characteristics when $1 < \gamma \leq 2$. When $\gamma=1$ and $\alpha=\beta=2$ we recover the classical diffusion equation. For $\gamma=1$ and $0 < [\alpha,\beta] \leq 2$, Eq.(\ref{FPDE}) describes a space-fractional diffusion process whose solutions belong to the superdiffusion regime \cite{meerschaert2004finite}. The superdiffusive behavior is the results of dynamics dominated by long-range interactions (the so called L{\'e}vy flights), where the step length distribution decays asymptotically as $x^{-(\alpha+1)}$.
This is also the interval of interest for the present study in which we want to explore the regime characterized by heavy-tailed distributions of the field quantities due to L{\'e}vy flights dominated dynamics.

In the following numerical analysis, we will consider the case of $\gamma=1 $ and $\alpha=\beta$. Once complemented with proper boundary conditions, this space-fractional diffusion equation (\ref{FPDE}) can be solved numerically. 

\begin{figure}[h!]
\centering
\includegraphics[width=9 cm]{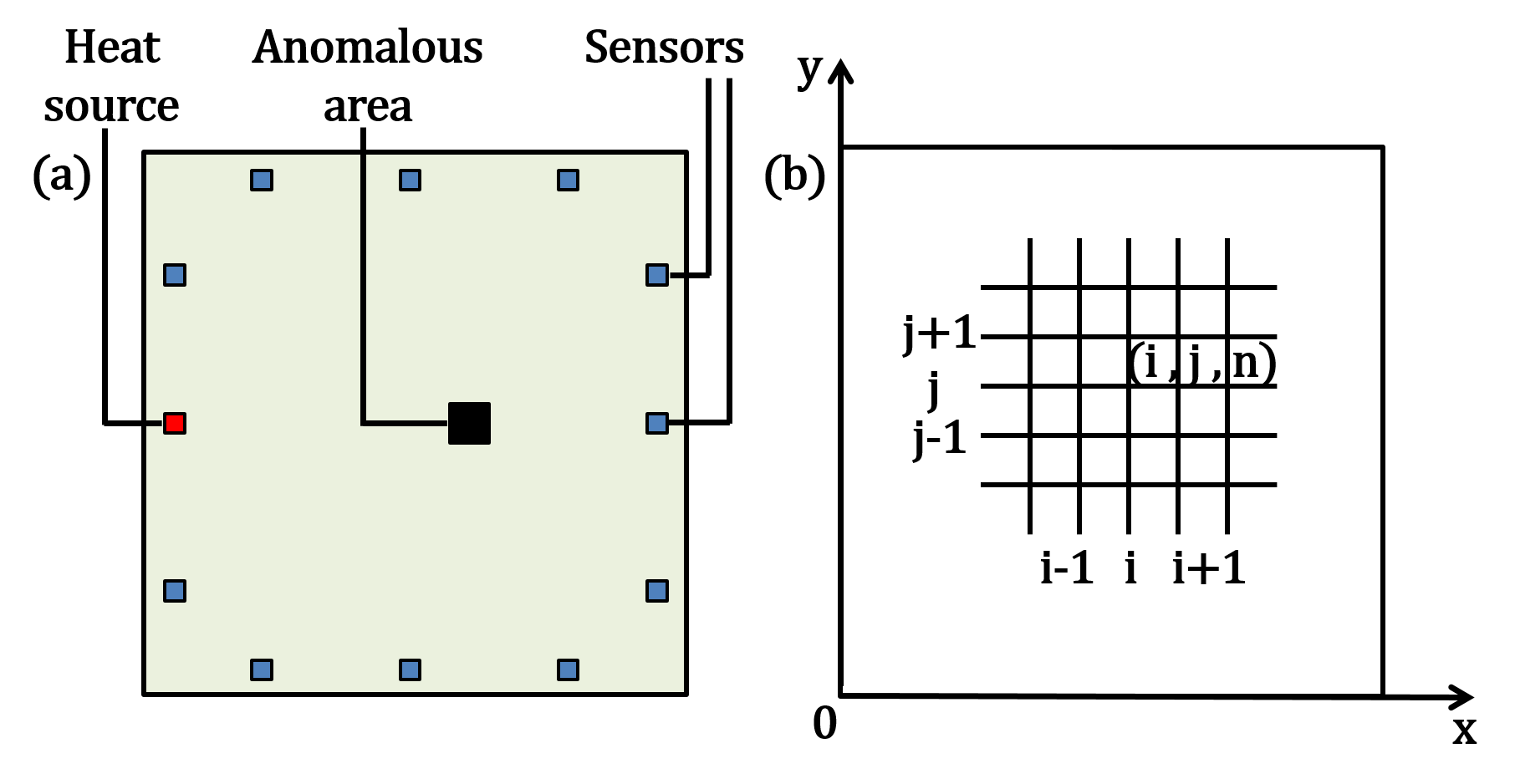}
\caption{(a) Schematics of the thermal tomographic computational setup indicating the location of the transducers used for both excitation and sensing. (b) The numerical uniform grid used to discretize the fractional diffusion equations.}
\label{Fig0}
\end{figure}

The thermographic problem considered below is solved using the following initial conditions:
\begin{eqnarray} \label{eqn:Initial condition}
T(x,y,t=0)=T_0(x,y)     \quad (x,y)\in \Omega \cup \partial\Omega
\end{eqnarray}

and boundary conditions:
\begin{eqnarray} \label{eqn:Boundary conditions}
T(x,y,t)=0 & & (x,y)\in \partial\Omega  \hspace{0.15 cm}   t>0,
\end{eqnarray}

where $\Omega$ and $\partial\Omega$ indicate the 2D domain and its boundary (Fig.~\ref{Fig0}(b)), respectively. 

The initial value problem defined by the set of equations (\ref{FPDE},\ref{eqn:Initial condition},\ref{eqn:Boundary conditions}) can be solved numerically using a shifted Grunwald-Letnikov (GL) finite difference (FD) scheme \cite{meerschaert2004finite}. More specifically, the two-sided shifted Grunwald-Letnikov scheme \cite{meerschaert2006finite} was selected due to its ability to respect the local symmetry of the solution at every node of the domain. Fig.~\ref{Fig0}(b) shows a uniform discretization grid used for the numerical solution of equation Eq.(\ref{FPDE}). The grid has a uniform spacing with $\Delta x=(b-a)/I $ and $\Delta y=(d-c)/J $, where $I$ and $J$ are the total number of discretization nodes in the $x$ and $y$ directions, respectively. The finite difference approximation to $T(x_i,y_j,t^n)$ will be denoted with $T_{ij}^n$ where $x_i=i\Delta x $, $y_j=j\Delta y$ and $t^n=n\Delta t $.

Following the above definition, the general form of the shifted Grunwald-Letnikov formulas are \cite{meerschaert2004finite}:
\begin{eqnarray} \label{eqn:Shifted Right and Left Handed Grunwald}
\begin{split}
\cfrac{d^{\alpha}f(x)}{d_+x^\alpha}
=\lim_{M_+\rightarrow \infty}\frac{1}{h^\alpha}\sum_{k=0}^{M_+}g_k\cdot f[x-(k-1)h]\\
\cfrac{d^{\alpha}f(x)}{d_-x^\alpha}
=\lim_{M_-\rightarrow \infty}\frac{1}{h^\alpha}\sum_{k=0}^{M_-}g_k\cdot f[x+(k-1)h]
\end{split}
\end{eqnarray}
where $M_+$, $M_-$ are positive integers, $h_+=(x-a)/M_+$, $h_-=(b-x)/M_-$ and \textbf{$k=1,2,3...M_{(\pm)}$}. The coefficients $g_k$ are the Grunwald weights expressed as:

\begin{eqnarray} \label{eqn:Grunwald weights}
 g_k=
 \begin{cases}
 1  &k=0\\    
 (-1)^k\frac{(\alpha)(\alpha-1)...(\alpha-k+1)}{k!}  \qquad &k=1,2,...M_{(\pm)}
 \end{cases}
\end{eqnarray}

Using the shifted GL relations (\ref{eqn:Shifted Right and Left Handed Grunwald}) to discretize Eq.(\ref{FPDE}), we obtain \cite{meerschaert2006finite}:

\begin{eqnarray} \label{eqn:Discretized 1}
\begin{split}
\frac{T_{ij}^{n+1}-T_{ij}^{n}}{\Delta t}=\frac{c_{ij}^n}{h^\alpha} \bigg[ \sum_{k=0}^{i+1}g_k\cdot T_{i-k+1}^n+\sum_{k=0}^{I-i+1}g_k\cdot T_{i+k-1}^n+  \\  \sum_{k=0}^{j+1}g_k\cdot T_{j-k+1}^n+\sum_{k=0}^{J-j+1}g_k\cdot T_{j+k-1}^n  \bigg]+s_{ij}^{n}
\end{split}
\end{eqnarray}

that can be explicitly solved for $T_{ij}^{n+1}$.

The stability condition of this scheme for the case $1<\alpha<2$ is given by \cite{meerschaert2006finite}: 
\begin{eqnarray} \label{Stability condition}
\beta = \frac{\Delta t}{h^\alpha} \leq \frac{1}{2\cdot\alpha\cdot c}
\end{eqnarray}

\subsubsection{Inverse problem}

The solution of the inverse problem is performed according to a classical iterative minimization approach following the Levenberg-Marquardt (LM) method. According to the LM method:
\begin{eqnarray} 
\label{eqn:LMM}
\boldsymbol{P^{ k+1}}=\boldsymbol{P^{k}}+\left [\mu^k\boldsymbol{\Omega^k}+(\boldsymbol{J^k})^T\boldsymbol{J^k}\right]^{-1}(\boldsymbol{J^k})^T[\boldsymbol{Y-T(P^k)}]
\end{eqnarray}

where $\boldsymbol{P}$ is the vector of design variables consisting in the system parameters (e.g. the coefficients $c$ or the fractional order $\alpha$), $\boldsymbol{T}$ is the vector of numerically estimated temperatures, $\boldsymbol{Y}$ is the vector of measured temperatures, $\mu^k$ is a regularization parameter \cite{holder2004electrical}, $\boldsymbol{\Omega^k}$ is a regularization matrix chosen in our simulations as the identity matrix, $\boldsymbol{J}$ is the Jacobian matrix, and the superscript $()^{\textit{T}}$ indicates the transpose. The term $\mu^k\boldsymbol\Omega^k$ is a regularization term used to damp the instabilities introduced by the ill-conditioned Hessian matrix $\boldsymbol{J^TJ}$ \cite{ozisik2000inverse}. The generic element of the Jacobian, or sensitivity matrix, is defined as:
\begin{eqnarray} 
\label{eqn:JACOBIAN}
J_{lm}=\frac{\partial T_l}{\partial P_m}
\end{eqnarray}

with $l=1,2,...,L$ and $m=1,2,...,M$, where $M$ is the total number of unknown parameters, $L$ is the total number of measurements ($L\geq M$), $T_l$ is the $l^{th}$ estimated temperature, and $P_m$ is the $m^{th}$ unknown parameter. 
The iterative problem described by Eq. (\ref{eqn:LMM}) is repeated until a prescribed error threshold is satisfied. In our simulations, the exit conditions was chosen to be a threshold value of the residual error $|\boldsymbol{Y}- \boldsymbol{T(P^k)}|<10^{-5}$.

\section{Problem description and numerical experiments}
In this section, we apply the methodology discussed above to reconstruct the spatial distribution of the thermal diffusivity of a 2D domain. The main difference, compared to traditional thermography, is that we let the domain have a non-homogeneous distribution of the order of the underlying differential operator. In other terms, we consider media in which the transport process is non-homogeneous and can result from a combination of standard and anomalous diffusion.
The following three scenarios will be considered: 1) the reference domain exhibits a fractional-order behavior with localized integer-order sub-domains, 2) the reference domain exhibits integer-order behavior with localized fractional-order sub-domains, 3) the reference domain exhibits integer-order behavior with local sub-domains having either different fractional order or diffusivity coefficients with respect to the background. The selection of these case of studies has been made to illustrate some specific characteristics of the fractional operator within the context of an inverse problem.

 \subsection{Case 1: domain with fractional-order behavior and integer-order inhomogeneities} \label{case1}
  
\begin{figure}[h!]
	\centering
	\includegraphics[width=9 cm]{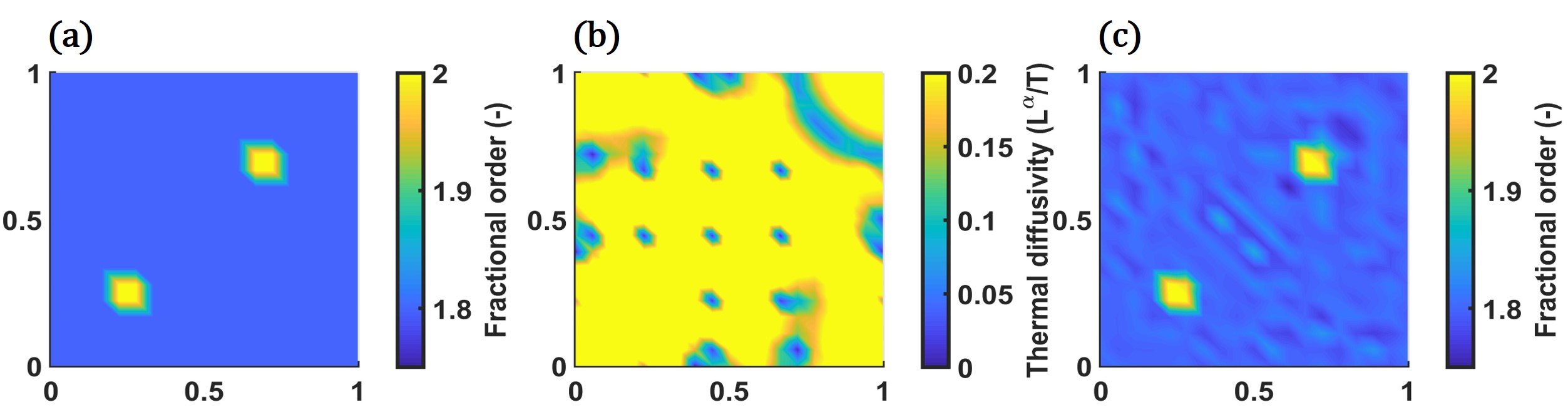}
	\caption{\textit{Case 1:} (a) Reference map of the 2D domain showing the distribution of the spatial order of the differential operator. (b) Reconstructed map of the thermal diffusivity distribution when an integer-order model is used for the forward problem. (c) Reconstructed map of the operator order when a fractional model is used for the forward problem. 
	}
	\label{Fig1}
\end{figure}

 This case addresses a situation in which the thermal field transport within the physical domain is described by a space-fractional diffusion equation. The domain includes two localized inhomogeneities consisting of areas where the transport is of integer order. The reference map showing the distribution of the order of the spatial differential operator is shown in Fig.~\ref{Fig1}$(a)$.  
 Note that this condition can be mapped to a physical situation where the domain of interest is characterized by long-range interactions (i.e. an anomalous diffusion process) other than for small localized areas that are controlled instead by short-range interaction (hence resulting in a standard diffusion process). 
  
  In order to illustrate the need for fractional order models in inverse problems involving anomalous transport behavior, we compared the results of the reconstruction obtained using both a fractional- and an integer-order forward model. In the case of the integer order model the only unknown parameter that can be reconstructed is the thermal diffusivity. The reconstructed spatial distribution of the thermal diffusivity $c(x,y)$ is shown in Fig.~\ref{Fig1}$(b)$. A direct visual inspection of the results indicates that the integer-order model is completely unable to capture the inhomogeneities in the domain due to the inability to simulate the underlying anomalous transport process. Conversely when a fractional-order model is used, the process successfully identifies (both in amplitude and location) the inhomogeneities present in the domain (Fig.~\ref{Fig1}$(c)$). In this case, the order of the fractional operator was selected as the unknown parameter describing the characteristics of the domain. 

The results above is likely the most significant among the three cases and clearly highlights the necessity of using a fractional-order model when the imaging is sought in domains characterized by anomalous field transport.
They also highlight another important feature of the fractional inverse problem. The fractional order can be used as an additional parameter to track both the presence, the magnitude, and the location of the inhomogeneities. In other terms, the fractional order expands the parameter space that can be used to characterize inhomogeneous domains (which could be relevant in applications such as non-intrusive sensing and non-destructive evaluation). 

The numerical experiments presented in the next two cases are intended to address and characterize the role of the fractional order as an additional indicator of inhomogeneities, that is beyond the traditional parameters such as the transport coefficients $c(x,y)$.

\subsection{Case 2: domain with integer-order behavior and fractional-order inhomogeneities}
This case explores the ability to image a domain characterized by local fractional-order inhomogeneities distributed in an otherwise integer-order background. Such situation physically corresponds to a domain in which the heat transfer in the background occurs according to a standard (Fourier) diffusion process but it experiences localized areas in which the transport is non-local.  
    
\begin{figure}[h!]
	\centering
	\includegraphics[width=9 cm]{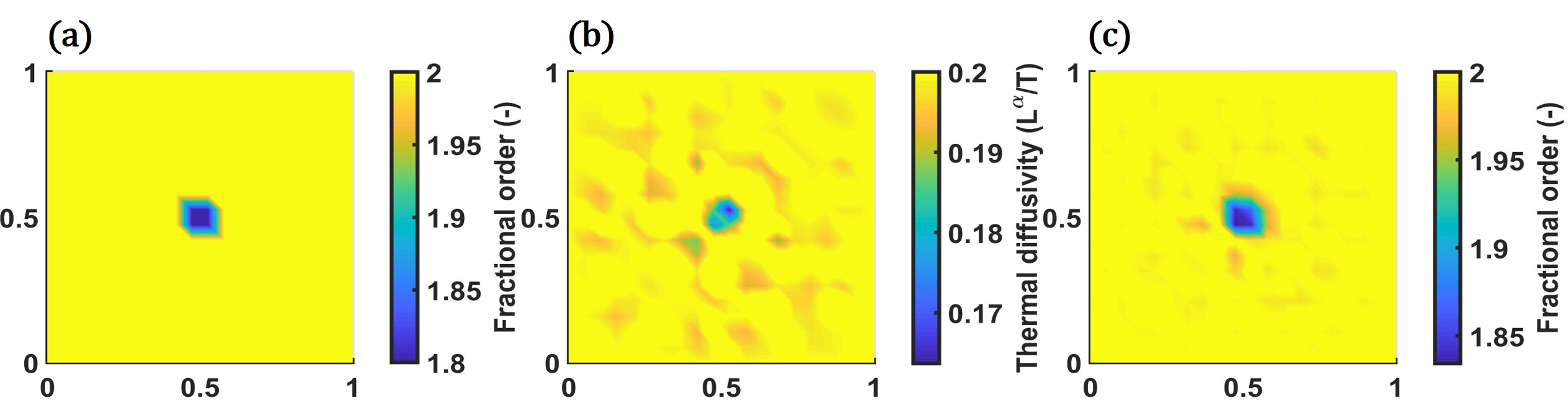}
	\caption{\textit{Case 2:} (a) Reference map for case 2 expressed in terms of the order of the operator. Reconstructed maps of (b) the diffusion coefficients and of (c) the fractional order. The inhomogeneity is represented by a discontinuity in the distribution of the fractional order $\alpha$ while the inverse procedure is set to reconstruct simultaneously the spatial distribution of the diffusion coefficients and of the fractional order.}
	\label{Fig2}
\end{figure}

The reference case to be reconstructed is shown in Fig.~\ref{Fig2}(a) where the background of the 2D domain has order $\alpha=2$ other than for a small area in the center where $\alpha=1.8$. The reconstruction was performed by using a fractional-order model where the thermal diffusivity coefficients and the fractional order were both treated as unknown variables.
  
To solve the inverse problem, the iterative approach was initialized using the true values of both $c$ and $\alpha$ in the background. The inverse procedure exited with a residual error of $5.44 e-10$ well below the set threshold.
Fig.~\ref{Fig2}(b) and Fig.~\ref{Fig2}(c) show the results of the reconstructed maps. In general, both parameter distributions show clear traces of the location of the inhomogeneity. However, a marked difference is visible in their corresponding amplitude. More specifically, the order of the differential operator is well captured by returning a value of $\alpha=1.835$ which is within $2\%$ error with respect to the reference. 

This was expected because the original inhomogeneity was set in terms of an anomaly in the order of the operator. The thermal diffusivity distribution (Fig.~\ref{Fig2}$(b)$) instead shows a variation of $16\%$ with respect to the background in correspondence to the inhomogeneity despite the original configuration was uniform in terms of $c(x,y)$. This result was not unexpected because previous studies \cite{John2017} had shown that there is a direct correspondence between spatially inhomogeneous coefficients of an integer order differential equation and the corresponding order of a matched fractional equation. We note that, although the location and extent of the anomaly is still well captured, the absolute value of the reconstructed coefficient is not necessarily meaningful and does not allow quantifying of the inhomogeneity.

\begin{figure}[h!]
\centering
\includegraphics[width=8 cm]{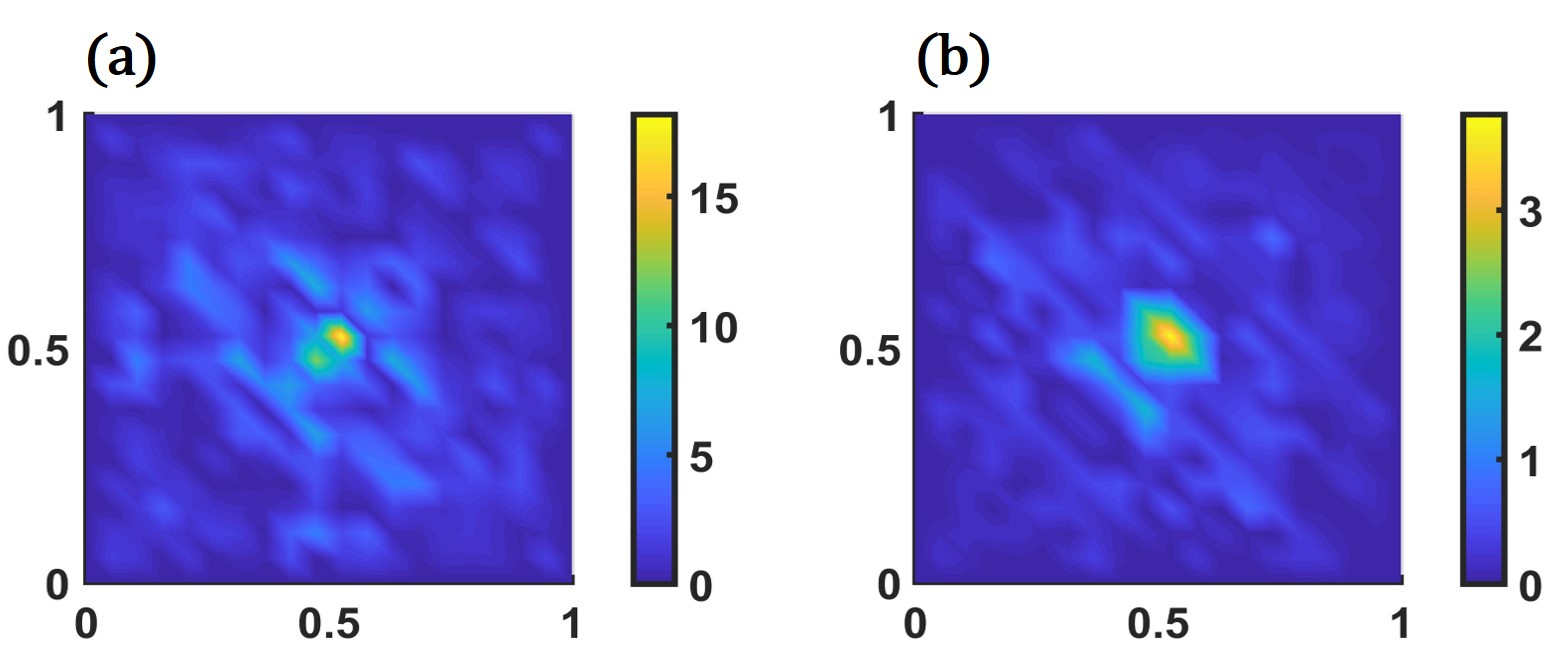}
\caption{\textit{Case 2}: Error maps. Percentage error maps of the (a) reconstructed thermal diffusivity and of the (b) reconstructed fractional order.}
\label{Fig3}
\end{figure}

\subsection{Case 3: domain with combined inhomogeneities of the order and diffusivity}

In this scenario, we explore the possibility to simultaneously reconstruct two distinct types of inhomogeneities affecting either the fractional order or the thermal diffusivity. This case is intended to underline the fundamental difference in the role played by both the thermo-physical properties (captured in the coefficients of the differential equation) and by the physical trasport mechanisms (captured by the fractional operator). In fact, variations of the thermal diffusivity do not affect the functional form of the solution of the heat conduction equation, while perturbations in the order of the operator do affect the transport process by introducing non-local effects. The underlying idea is that a change in the nonlocal parameter $\alpha$ indicates the occurrence of non-local heat transfer which ultimately is connected to changes in the properties of the host medium.

\begin{figure}[h!]
	\centering
	\includegraphics[width=9 cm]{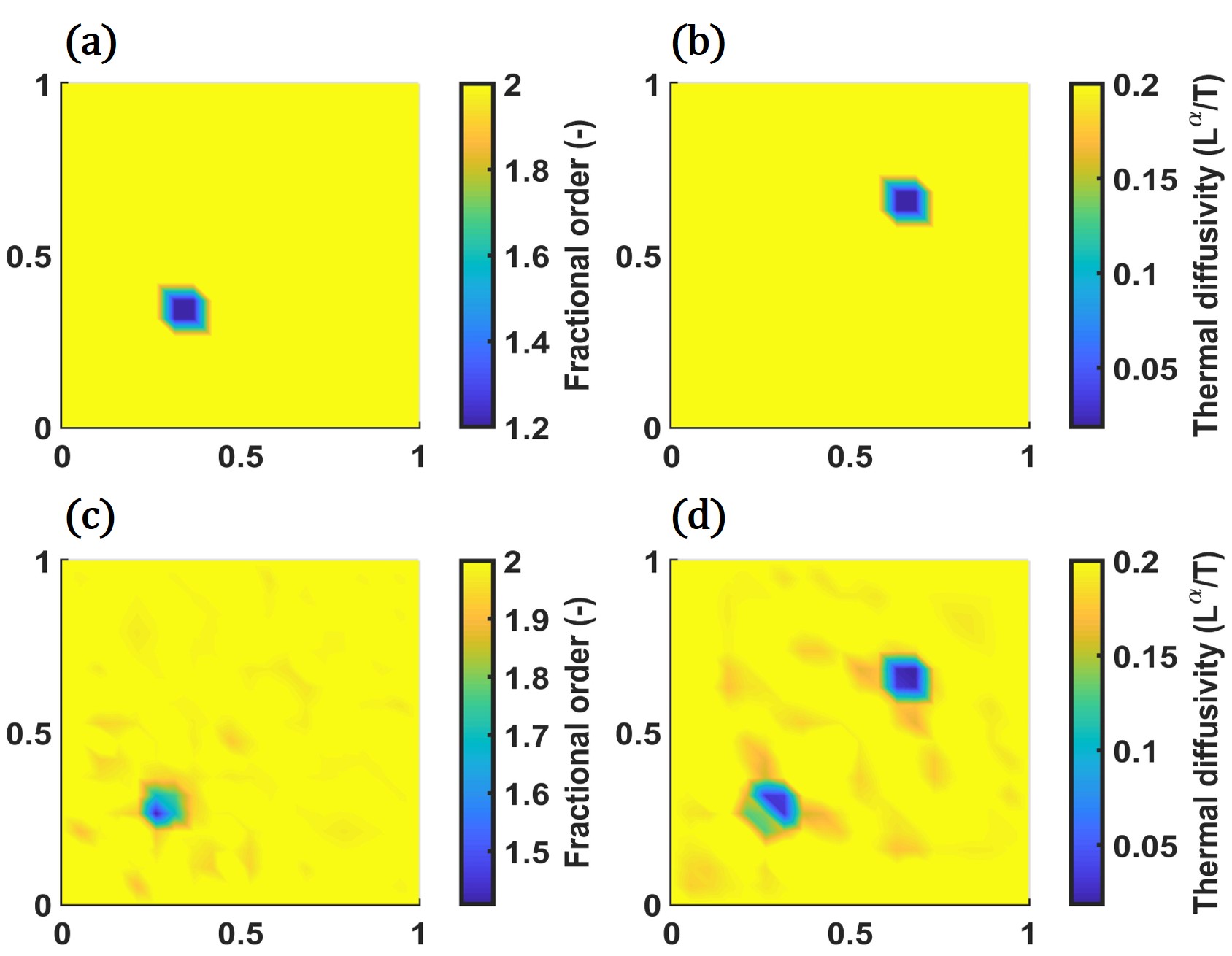}
	\caption{\textit{Case 3:} (a-b) Reference and reconstructed maps of the (c) fractional order and of the (d) diffusion coefficients. The damages are modeled by a discontinuity in the distribution of fractional order $\alpha$ and a discontinuity in the distribution of the thermal diffusivity $c$. The inverse procedure is set to reconstruct simultaneously the spatial distribution of the diffusion coefficients and of the fractional order.}
	\label{Fig8}
\end{figure}

Fig.~\ref{Fig8} summarizes the results of the numerical reconstruction. The background values of $c$ and $\alpha$ are the same of the previous case, and the two inhomogeneities are represented by an anomaly in the thermal diffusivity of intensity $c = 0.02$ \textbf{$[L^\alpha/t]$} (Fig.~\ref{Fig8}$(b)$) and an anomaly in the fractional order distribution equal to $\alpha=1.2$ (Fig.~\ref{Fig8}$(a)$). The iterative inverse procedure was initialized using as initial guess the true values of the background distributions of $c$ and $\alpha$. The inverse procedure exited with a residual error of $1e-5$.

\begin{figure}[h!]
\centering
\includegraphics[width=8 cm]{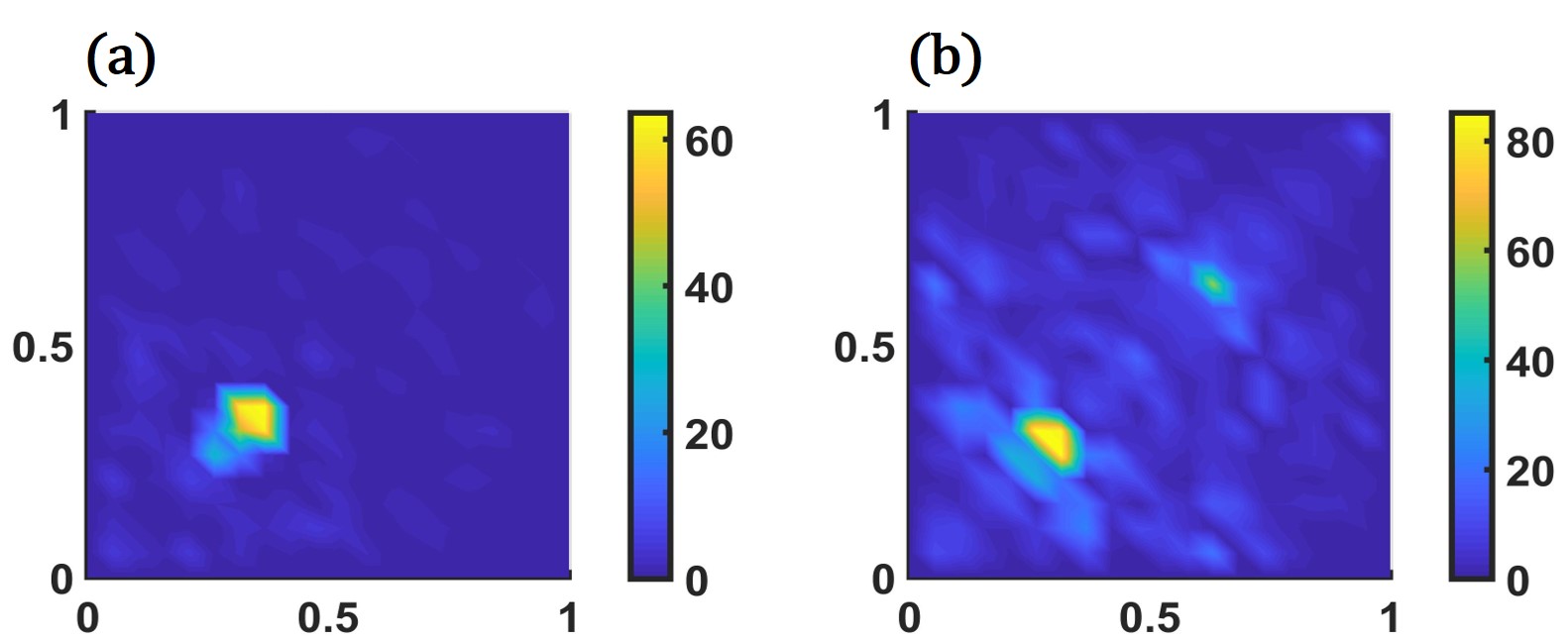}
\caption{\textit{Case 3:} Error maps. Percentage error maps of the (a) reconstructed fractional order and of the (b) reconstructed thermal diffusivity.}
\label{Fig9}
\end{figure}

The reconstructed distributions could correctly identify the location and intensity of both the inhomogeneities in terms of
fractional order (Fig.~\ref{Fig8}$(c)$) and thermal diffusivity (Fig.~\ref{Fig8}$(d)$). A marked inhomogeneity emerged in the reconstructed map of the diffusion coefficients. This result should have been expected given that, as shown in case 2, there is a strict correspondence between the spatially inhomogeneous coefficients of an integer order differential equation and the corresponding order of a matched fractional equation. This link between the system parameters also tend to reduce the accuracy of the identification of the fractional order, as seen in Fig.~\ref{Fig8}$(c)$. The indeterminacy that occurs when both parameters are perturbed cannot be resolved mathematically unless more information on the host system was provided. As an example, in cases where \textit{a priori} information on the inhomogeneity was available, then the inverse problem could be solved using constraints on the optimization problem so the facilitate resolving indeterminacy. Such situations might correspond to cases in which, as an example, a knowledge of the materials involved or of possible structural defects was available from other measurements.
Fig.~\ref{Fig9}$(a)-(b)$ shows the percentage error maps for the thermal diffusivity and the fractional order.

The results of the numerical experiments suggest that the augmented parameter space resulting from the use of fractional order models has the potential to improve significantly the accuracy of the reconstruction procedure when compared to the conventional tomographic techniques. 

In fact, the nonlocal parameter introduced by the fractional formulation (i.e. the order of the operator) serves as a new design variable able to capture the underlying nature of the heat conduction process as well as of describing the anomalous behavior observed in complex and inhomogeneous media. In addition, fractional tomography opens the way to exploit hybrid field transport (i.e. a combination of propagating and diffusive mechanisms) as a probing mechanism, whereas classical tomography is constrained to the use of a single transport process.

Overall, the above results suggest that the enrichment of the parameter space due to the integration of fractional order models in tomographic methods, may have important effects on the sensitivity of the reconstruction process with respect to classical tomography. 

\section{Conclusions}   \label{Conclusions}

This study presented the general theoretical and mathematical framework for the implementation of fractional tomographic imaging. This general class of techniques was conceived to achieve non-intrusive and non-destructive imaging in a variety of applications where the host medium is characterized by anomalous and hybrid field transport. Among these applications, we can include imaging of viscous, nonlocal, and highly scattering media.

The proposed fractional tomographic method was tested via numerical experiments. In this study, the fractional nature of the model was limited to the spatial derivative in order to target specifically nonlocal and anomalous diffusion processes. Numerical results showed that the order of the operator can be used as a sensitive and effective parameter to probe and image the interior of a medium. This approach provides, not only a tomographic framework with an augmented parameter space, but also a powerful method for sensing and imaging in nonlocal complex media that traditionally pose extremely challenging conditions. Even though further research is required to establish constitutive relations able to connect the inhomogeneity of the system parameters to the fractional order, the present study shows clear evidence of the potential of the fractional tomographic framework and its ability to considerably enhance sensitivity and resolution in remote sensing applications.

\section{Acknowledgments}
The authors gratefully acknowledge the financial support of the National Science Foundation under the grant CMMI CAREER $\#1621909$.

\bibliographystyle{apsrev}
\bibliography{aipsamp}

\end{document}